\documentclass{PoS}

\title{Comparative study of topological charge}

\ShortTitle{Comparative study of topological charge}

\author{\speaker{Yusuke Namekawa}\\
        Center for Computational Sciences, University of Tsukuba,
        Tsukuba, Ibaraki 305-8577, Japan\\
        E-mail: \email{namekawa@ccs.tsukuba.ac.jp}}

\abstract{
Comparative study of topological charge is performed.
Topological charges are measured by a cloverleaf operator
on smoothed gauge configurations.
Various types of smoothing techniques are employed.
Agreement of topological charges in fermionic and gluonic definitions
is examined.
High consistency is observed between topological charges obtained by improved smoothing methods
and those by the index theorem with the overlap-Dirac operator.
}

\FullConference{
 The 32nd International Symposium on Lattice Field Theory,\\
 23-28 June, 2014\\
 Columbia University New York, NY
}

\begin{document}

\section{Introduction}

A topological charge is one of the most fundamental quantity in QCD.
It characterizes the vacuum structure.
Lattice QCD is a main tool in the study of
the topological charge~\cite{Plenary_2014}.
Lattice QCD allows us to perform a nonperturbative analysis
in a systematic way.

The topological charge is often measured
with a gluonic field strength operator
on the lattice.
Though it suffers from noisy ultraviolet fluctuations,
a smoothing technique tames them
so that a discernible signal of the topological charge can be obtained.
Cooling or smearing have been used for smoothing gauge fields.
Recently, a gradient flow is also employed.
In contrast to the traditional cooling and smearing,
the gradient flow has an advantage that it provides
a continuous change of the gauge field.
The gradient flow accomplishes a better control of smoothing.

Alternatively,
the topological charge can be calculated with a fermionic definition.
%
%
The topological charge is determined, for example,
by the index theorem with the overlap-Dirac operator.
A clear advantage of the fermionic definition is that
the result is guaranteed to be an integer.
A subtle point, on the other hand, is
the integer values depend on the choice of the definition,
due to a finite lattice spacing.
In the case of the overlap-Dirac operator,
a value of the topological charge occasionally changes according to
a parameter in the formulation.
%
Consistency check of the topological charges in the fermionic and gluonic definitions
would be helpful
as an estimator of the scaling violation.

In this work,
topological charges are computed with gluonic operators
on $N_f=2$ topology fixed gauge configurations.
%
The measurements are performed using several smoothing techniques.
Cooling with plaquette and improved local actions, APE and HYP smearing,
as well as gradient flows are employed.
The results are compared with each other,
and with the values obtained using the overlap-Dirac operator.
%
Similar attempts are reported in Refs.~\cite{Talk_2014}.

\section{Setup}

\subsection{Gauge configuration}

Measurement of the topological charge is performed
on $N_f=2$ gauge configurations provided by JLQCD Collaboration~\cite{JLQCD_2008}.
The lattice size is $16^3 \times 32$ at the lattice spacing of $a = 0.118(2)$ fm.
The gluon action is Iwasaki-type improved gauge action,
\begin{eqnarray}
 S_g
 = \beta
   \left(
     c_0^g \sum_{x,\mu < \nu} P_{\mu\nu}(x)
    +c_1^g \sum_{x,\mu,  \nu} R_{\mu\nu}(x)
   \right),
\end{eqnarray}
where $\beta = 6 / g_0^2$, $c_0^g = 3.648$, $c_1^g = -0.331$.
The quark action is an overlap-Dirac fermion action,
\begin{eqnarray}
 S_q
 &=& \bar{q} D_{\rm ov}(m) q,
 \\
 D_{\rm ov}(m)
 &=& \left( m_0 + \frac{m}{2} \right)
    +\left( m_0 - \frac{m}{2} \right) \gamma_5 \mbox{sgn}(H_{\rm W}(-m_0)),
 %
 H_{\rm W}(-m_0)
 =
 \gamma_5 D_{\rm W}(-m_0),
\end{eqnarray}
where $m$ is the bare quark mass.
$D_{\rm W}(-m_0)$ is the Wilson operator
with a negative mass, $-m_0 = -1.6$.
Furthermore,
unphysical Wilson fermion $\psi$ with a negative mass
as well as twisted mass terms
are added to fix the topological charge defined by the index theorem $Q_{\rm index}$,
\begin{eqnarray}
 \delta S_{\rm W}
 =
 \bar{\psi} D_{\rm W}(-m_0) \psi + \phi^{\dagger} (D_{\rm W}(-m_0) + i \mu \gamma_5 \tau_3) \phi,
\end{eqnarray}
where $\phi$ is a pseudofermion, and $\mu$ is the twisted mass parameter.
$\mu = 0.2$ is employed
in the configuration generation.
%
Each configuration is separated by 100 trajectories with its trajectory length 0.5.
The main simulation parameters are summarized in Table~\ref{table:simulation_parameter}.

\begin{table}[t]
\begin{center}
\begin{tabular}{ccccc}
\hline
 $\beta$           & $m$        & $Q_{\rm index}$ & 
 \# conf           & MD time
\\ \hline
 2.30              & 0.05       & -2              &
  50               & 2500
\\ \hline
%
%
\end{tabular}
\caption{Parameters of the gauge configurations.
         Molecular Dynamics time is the number of trajectories
         multiplied by the trajectory length.
}
\label{table:simulation_parameter}
\end{center}
\end{table}

\subsection{Gluonic topological charge operator}

\begin{figure}[t]
\begin{center}
 \includegraphics[width=75mm]{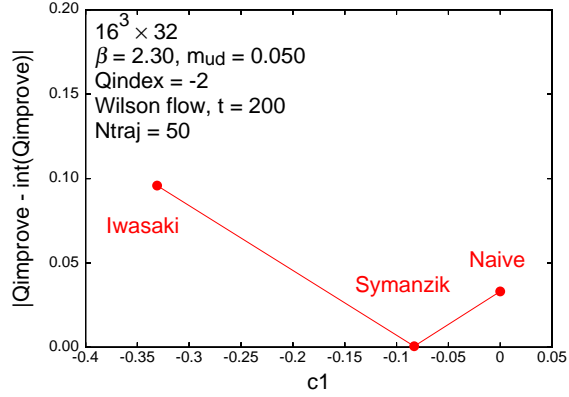}
 \caption{
   Improvement coefficient $c_1$ dependence of improved topological charge $Q_{\rm improve}$
   on a single gauge configuration.
   Difference of $Q_{\rm improve}$ from an integer is plotted.
 }
 \label{figure:Q_improve_vs_c_1}
\end{center}
\end{figure}

The topological charge is measured
using a gluonic field strength.
\begin{eqnarray}
 Q_{\rm improve}
 &=&
 c_0 Q^{1 \times 1} + c_1 Q^{1 \times 2},
 \label{equation:topological_charge}
 \\
 Q^{1 \times 1,2}
 &=& \frac{1}{32 \pi^2}
      \sum_{\mu,\nu,\rho,\sigma}
      \epsilon_{\mu \nu \rho \sigma}
      \mbox{Tr } F_{\mu \nu}^{1 \times 1,2} F_{\rho \sigma}^{1 \times 1,2},
 \\
 F_{\mu \nu}^{1 \times 1,2}
 &=&
 - \frac{i}{4} [C_{\mu \nu}^{1 \times 1,2}]^{\rm AH},
 %
 C_{\mu\nu}^{\rm AH}
 =
 \frac{1}{2 i} \left( C_{\mu\nu} - C_{\mu\nu}^{\dagger} \right),
\end{eqnarray}
where $C_{\mu \nu}^{1 \times 1}$ is the cloverleaf
constructed with $1 \times 1$ plaquettes,
and $C_{\mu \nu}^{1 \times 2}$ with $1 \times 2$ rectangular loops.
The improvement coefficients $c_0$ and $c_1$ can be tuned
to reduce the scaling violation in the topological charge operator.
Three types of $(c_0, c_1)$ are investigated.
\begin{eqnarray}
 (c_0, c_1)
 &=& (1, 0)          \mbox{\ \ \ Naive-type},
 \label{equation:naive_action}
 \\
 &=& (5/3, -1/12)    \mbox{\ \ \ Symanzik-type},
 \\
 &=& (3.648, -0.331) \mbox{\ \ \ Iwasaki-type}.
 \label{equation:Iwasaki_action}
\end{eqnarray}
Figure~\ref{figure:Q_improve_vs_c_1} displays $c_1$ dependence of the topological charge
on a single configuration smoothed by Wilson flow.
The topological charge with Symanzik-type coefficients
has the smallest deviation from an integer.
Since the deviation is originated from the finite lattice spacing,
it implies an efficient reduction of the scaling violation
by Symanzik-type operator.
Calculations on other configurations show a similar tendency.
Based on this result, Symanzik-type coefficients are employed
in this work.

\subsection{Smoothing}

Three kinds of smoothing techniques are evaluated:
cooling, smearing, and gradient flow.
Smoothing is required to suppress noisy ultraviolet
fluctuations, while keeping a topological structure.
%
%
%
%
Although any smoothing is expected to give a consistent result in the continuum limit,
it is valuable to find a method that has the least lattice artifact.

Cooling eliminates ultraviolet noises
by replacing each link variable such that
the local action is minimized~\cite{Cooling_1981}.
For the local action, not only a naive plaquette action,
but also Symanzik and Iwasaki actions are employed
with the coefficients of Eq.~(\ref{equation:naive_action})--(\ref{equation:Iwasaki_action}).
%

Another way of smoothing is smearing.
APE smearing~\cite{APE_1987} is defined by
\begin{eqnarray}
 U_{\mu}^{\rm new}(x)
 &=& \mbox{Proj}_{\rm SU(3)}
     \left[ (1 - \alpha) U_{\mu}(x)
           +\frac{\alpha}{6} \Sigma_{\mu}(x)
     \right],
 \alpha = 0.6.
 \\
 \Sigma_{\mu}(x)
 &=& \sum_{\pm \nu \neq \mu}
     U_{\nu}(x) U_{\mu}(x+\nu) U_{\nu}^{\dagger}(x+\mu)
\end{eqnarray}
In addition,
HYP smearing~\cite{HYP_2001} is also examined.
\begin{eqnarray}
 U_{\mu}^{\rm new}(x)
 &=& \mbox{Proj}_{\rm SU(3)}
     \left[
       (1 - \alpha_1) U_{\mu}(x)
      +\frac{\alpha_1}{6} \sum_{\pm \nu \neq \mu}
                          U_{\nu;\mu}^{(2)}(x) U_{\mu;\nu}^{(2)}(x+\nu) U_{\nu,\mu}^{(2),\dagger}(x+\mu)
     \right],
 \\
 U_{\mu;\nu}^{(2)}(x)
 &=& \mbox{Proj}_{\rm SU(3)}
     \left[
       (1 - \alpha_2) U_{\mu}(x)
      +\frac{\alpha_2}{4} \sum_{\pm \rho \neq \mu,\nu}
                          U_{\rho;\mu\nu}^{(3)}(x) U_{\mu;\nu\rho}^{(3)}(x+\rho) U_{\rho,\mu\nu}^{(3),\dagger}(x+\mu)
     \right],
 \\
 U_{\mu;\nu\rho}^{(3)}(x)
 &=& \mbox{Proj}_{\rm SU(3)}
     \left[
       (1 - \alpha_3) U_{\mu}(x)
      +\frac{\alpha_3}{2} \sum_{\pm \sigma \neq \mu,\nu,\rho}
                          U_{\sigma}(x) U_{\mu}(x+\sigma) U_{\sigma}^{\dagger}(x+\mu)
     \right],
 \\
 \alpha_1
 &=& 0.75, \alpha_2=0.6, \alpha_3 = 0.3.
\end{eqnarray}
Smeared gauge fields are projected back to $SU(3)$ by Maximum $SU(3)$ projection.
\begin{eqnarray}
 \mbox{Proj}_{\rm SU(3)}^{\rm max SU(3)}(U_{\mu}(x))
 &=& \max_{U_{\mu}^{\rm new}(x) \in SU(3)} \mbox{Re Tr }(U_{\mu}^{\rm new}(x) U_{\mu}^{\dagger}(x)).
\end{eqnarray}

An alternative smoothing is given by the gradient flow~\cite{Luescher_2010}.
The evolution of the gauge field is determined by
\begin{eqnarray}
 \partial_t V_{\mu}(x,t)
 &=& - V_{\mu}(x,t) \frac{\partial S}{\partial V_{\mu}(x)},
 \\
 V_{\mu}(x,t=0)
 &=& U_{\mu}(x),
\end{eqnarray}
where $t$ is the flow time, and $S$ is an action without its coupling constant.
Similar to the cooling case,
plaquette, Symanzik, and Iwasaki actions are employed.
The flow equation is solved by the fourth order 
Runge-Kutta in the commutator-free method~\cite{Celledoni_2003}.
The Runge-Kutta step size $dt$ is chosen to be $0.02$.
%
%
The systematic error associated with discretization of the flow time
is definitely below the statistical error.


\subsection{Results}

\begin{figure}[t]
\begin{center}
 \includegraphics[width=45mm]{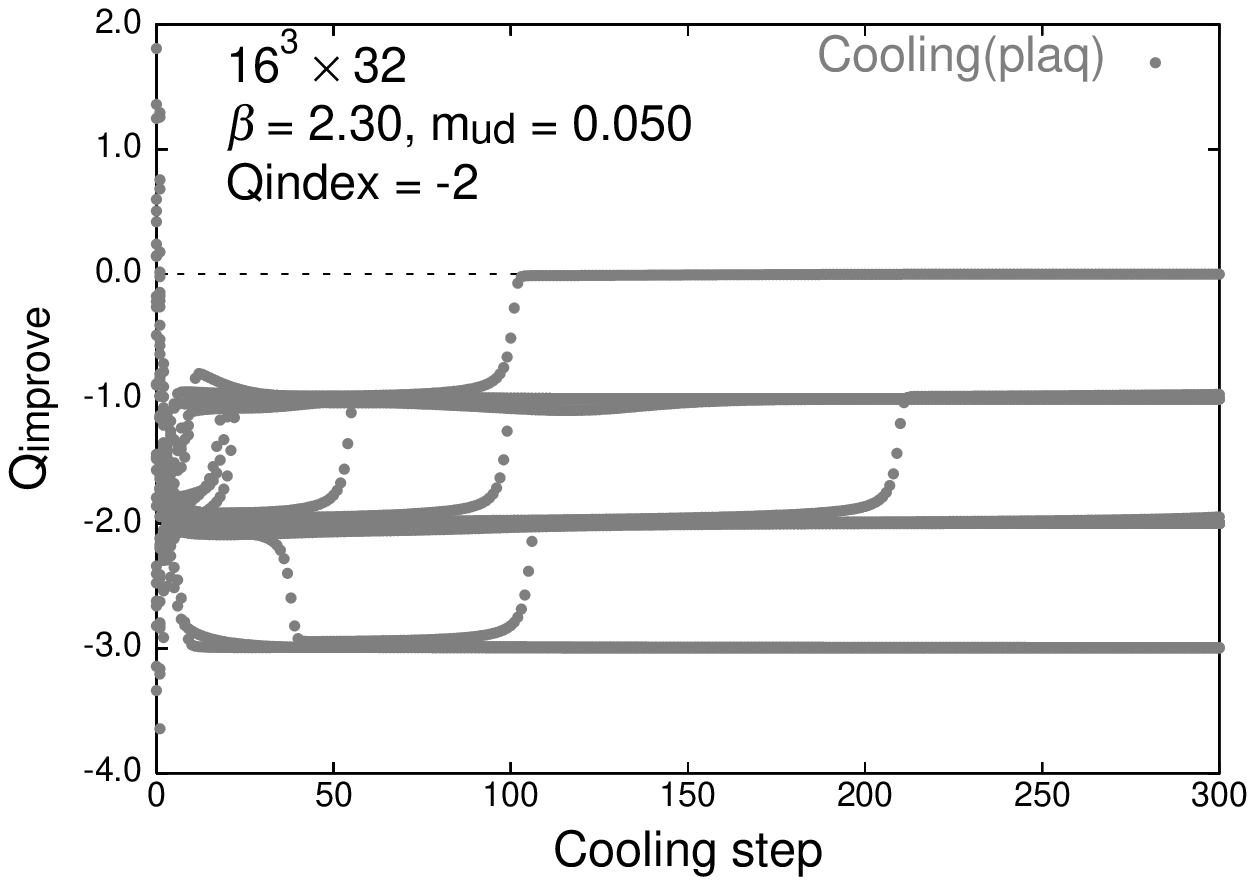}
 \includegraphics[width=45mm]{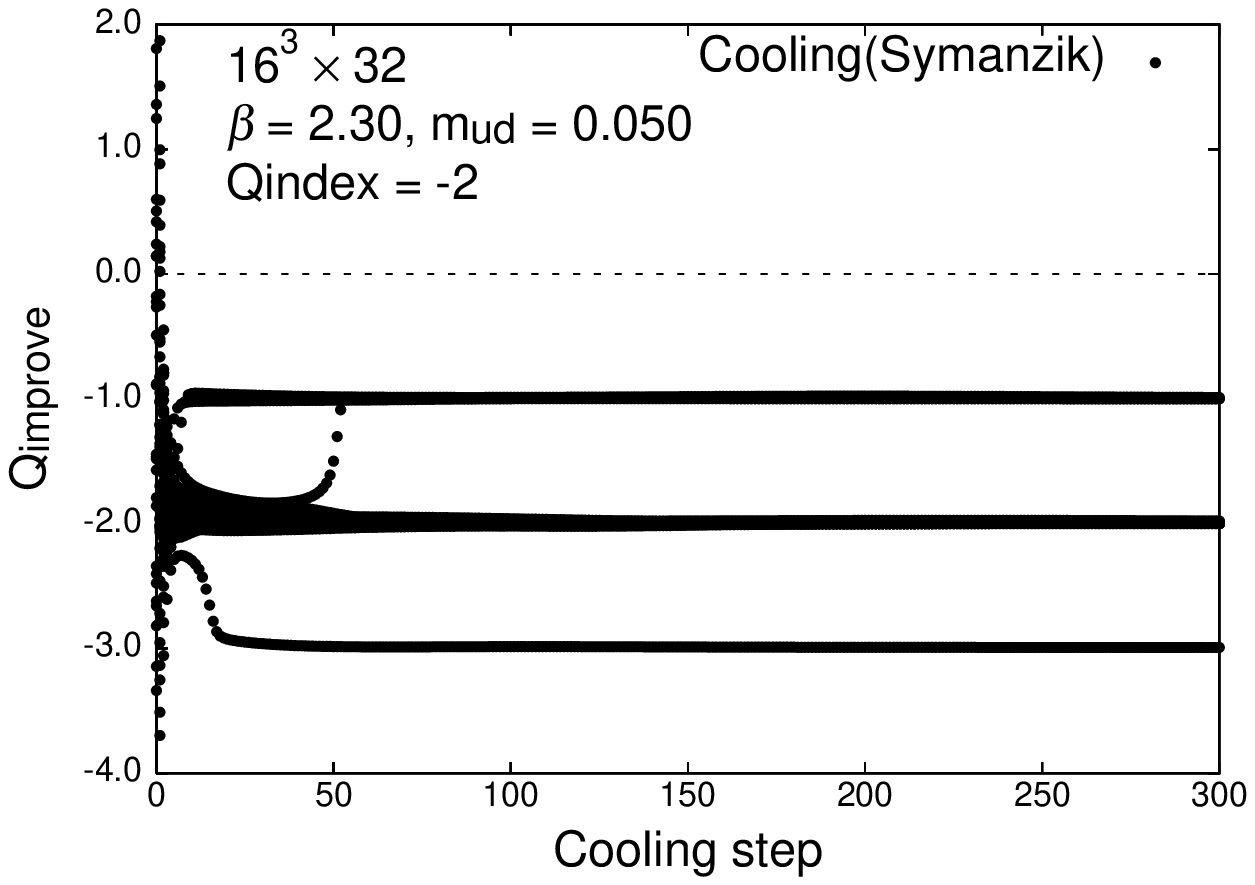}
 \includegraphics[width=45mm]{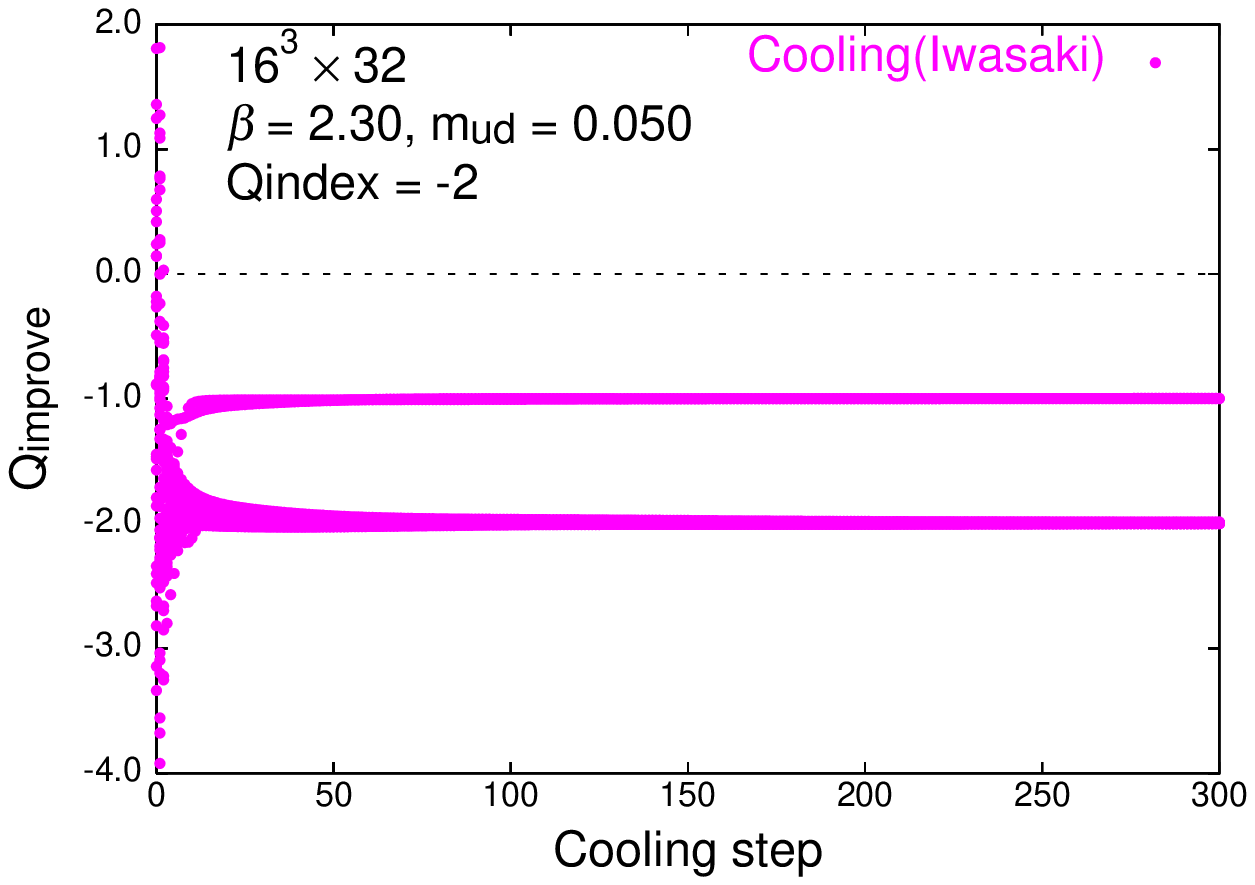}
 \\
 \includegraphics[width=45mm]{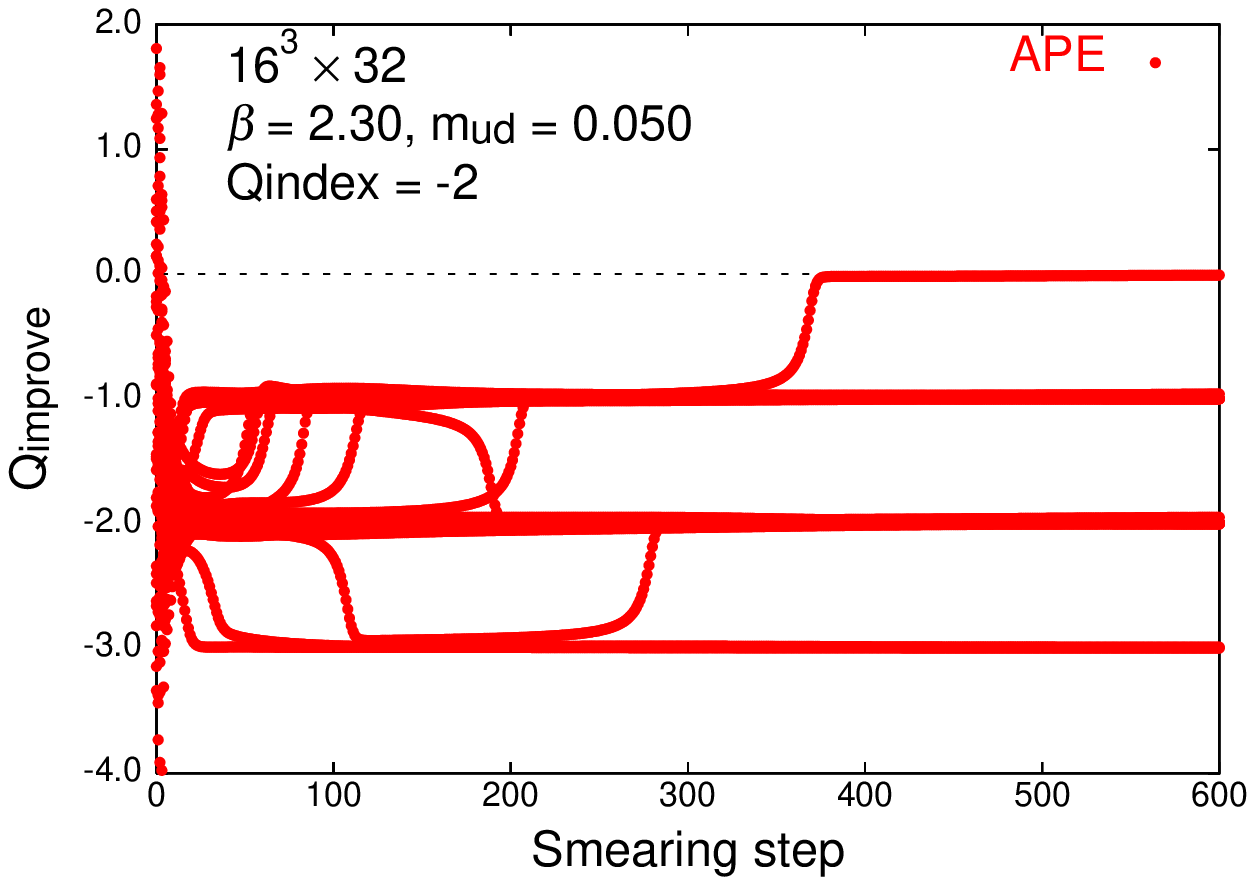}
 \includegraphics[width=45mm]{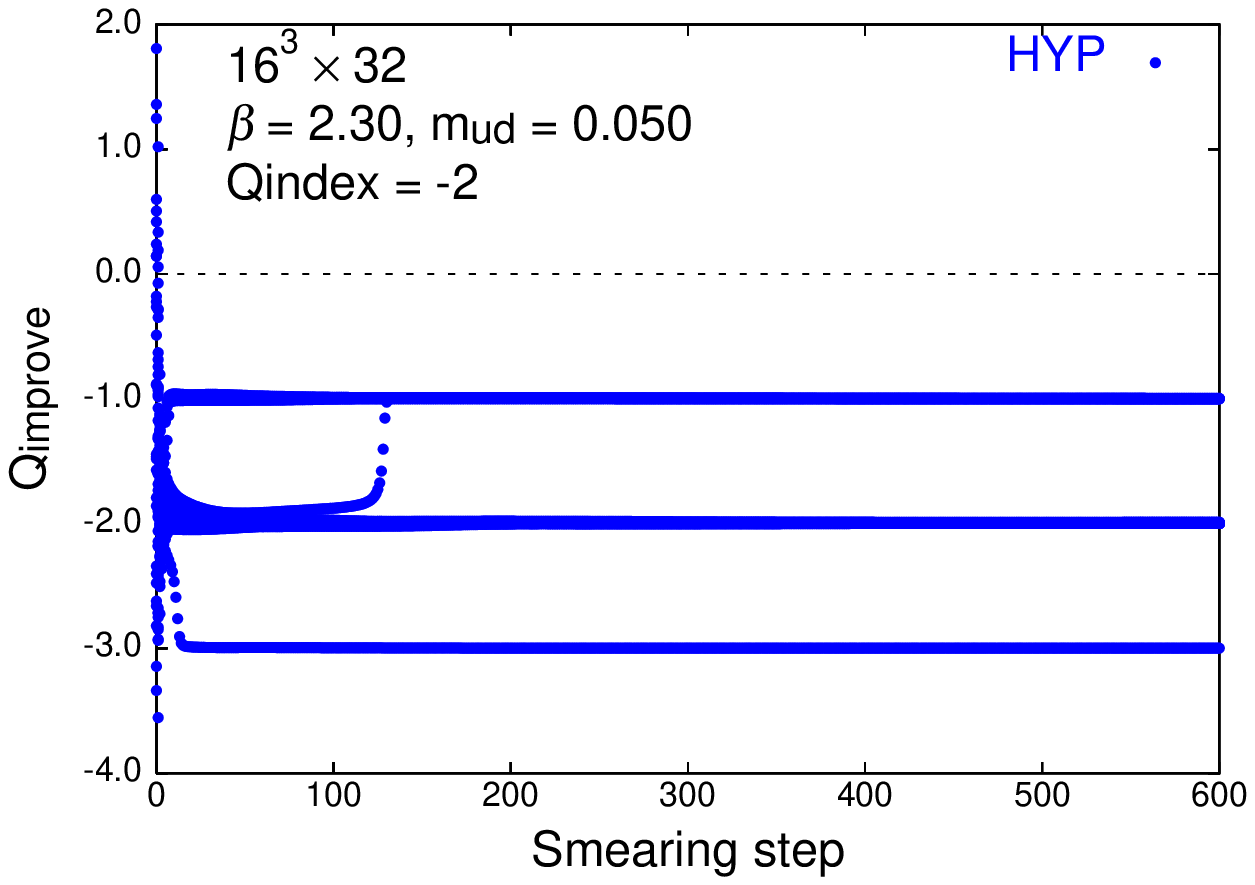}
 \\
 \includegraphics[width=45mm]{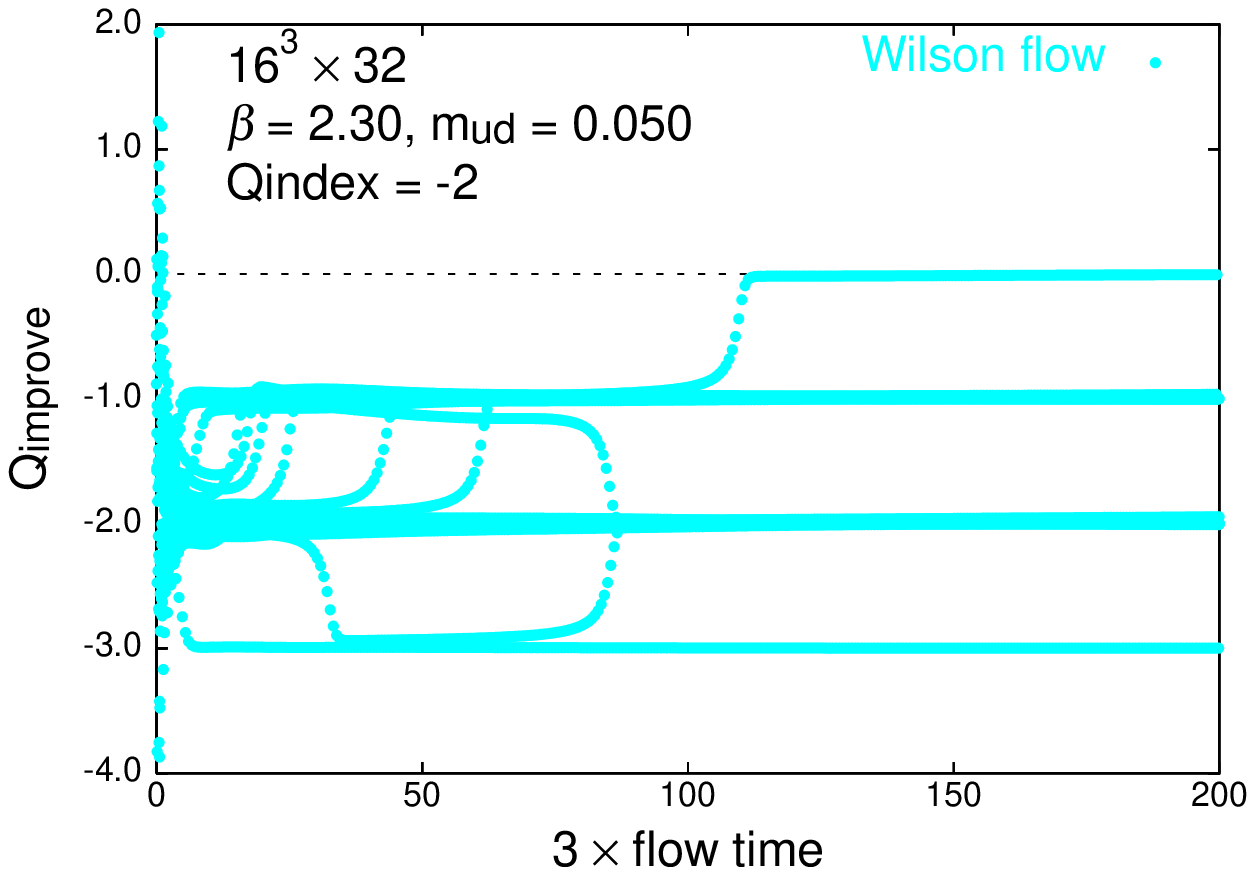}
 \includegraphics[width=45mm]{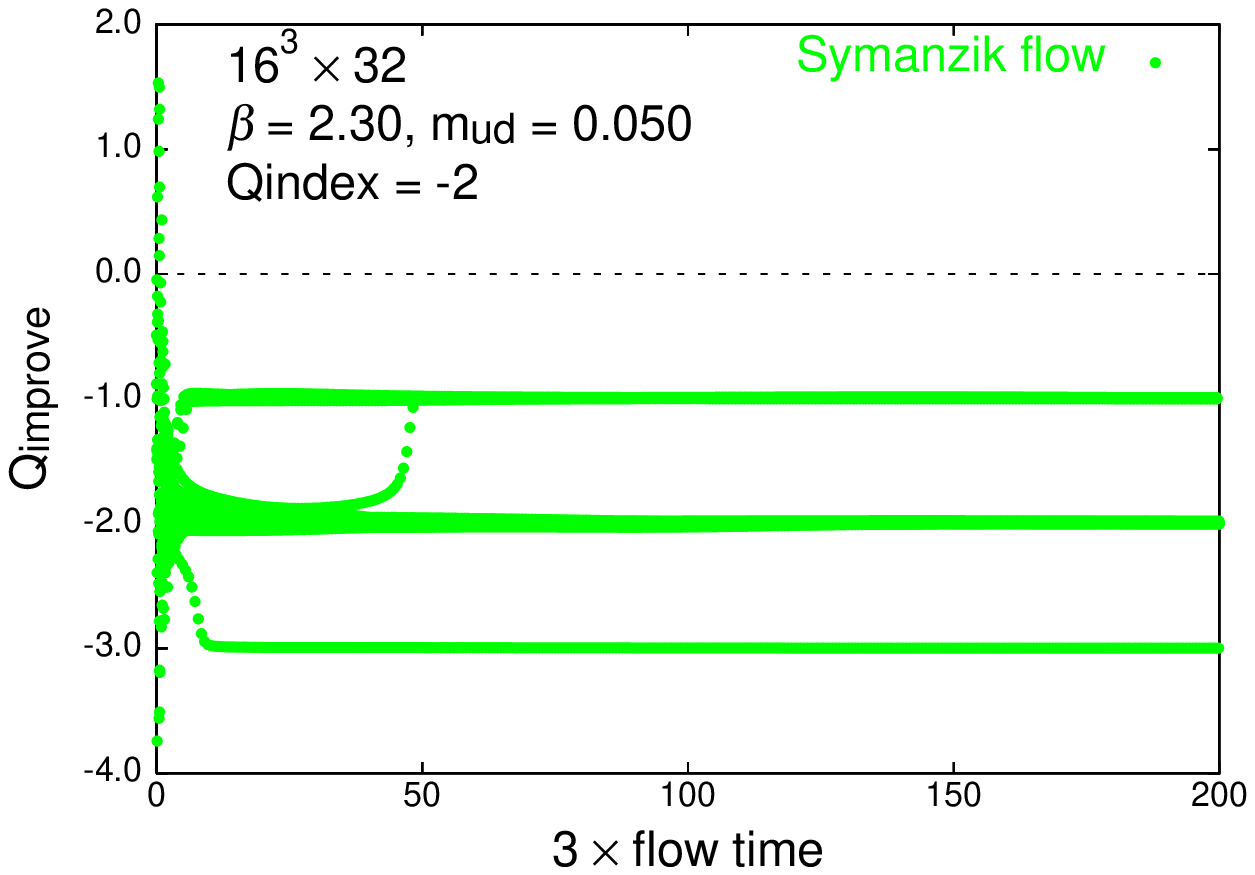}
 \includegraphics[width=45mm]{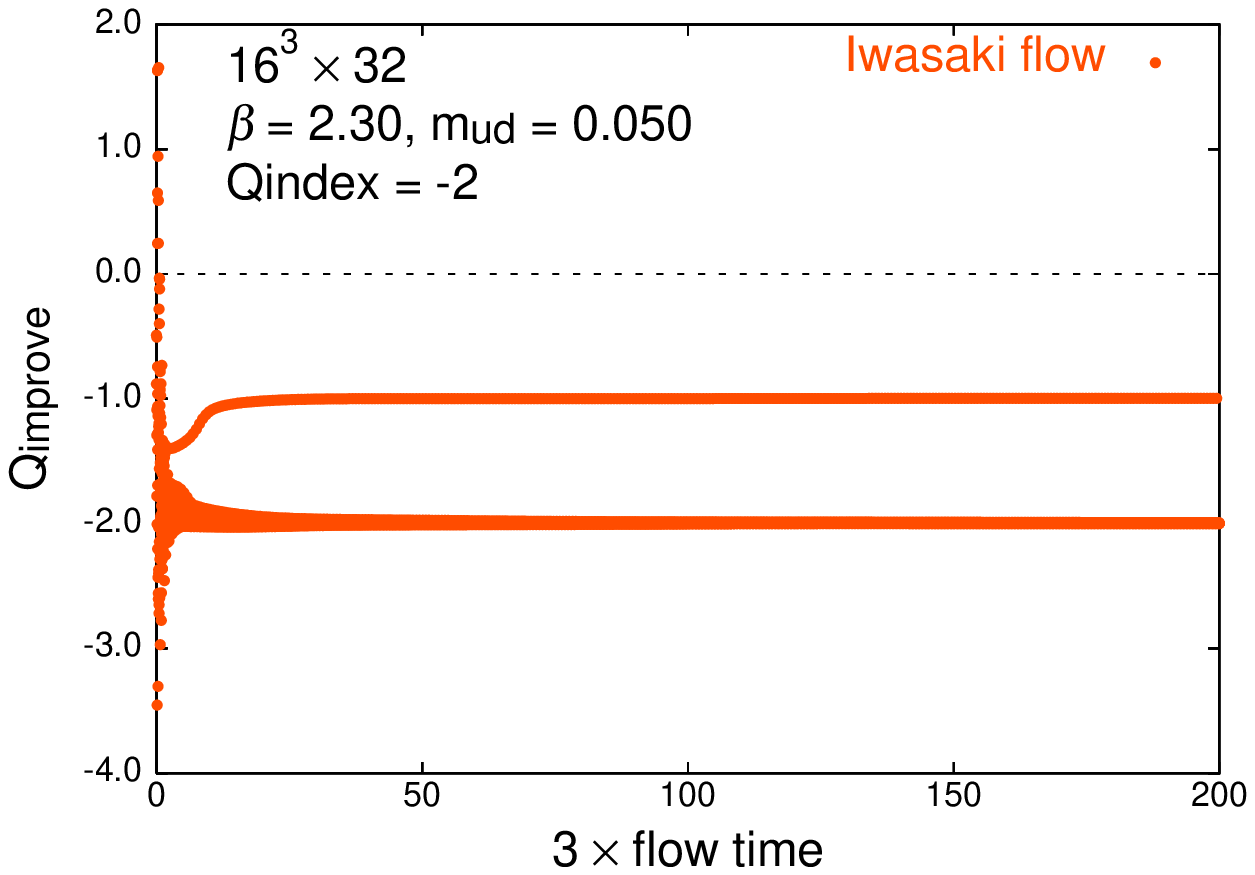}
 \caption{
   Smoothing step dependence of $Q_{\rm improve}$
   using cooling (top panels),
   smearing (middle panels),
   and gradient flow (bottom panels).
 }
 \label{figure:cooling_vs_topological_charge}
\end{center}
\end{figure}

\begin{figure}[t]
\begin{center}
 \includegraphics[width=75mm]{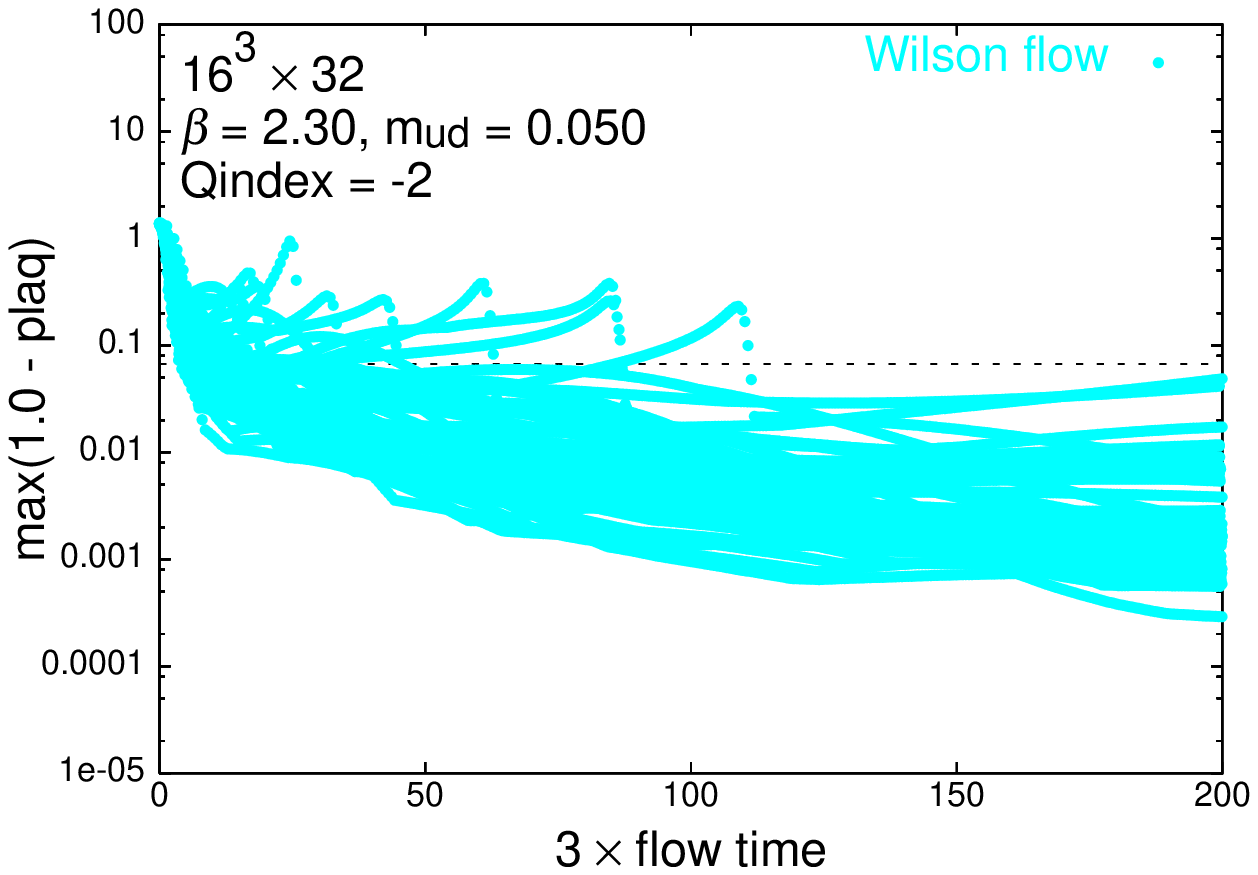}
 \includegraphics[width=75mm]{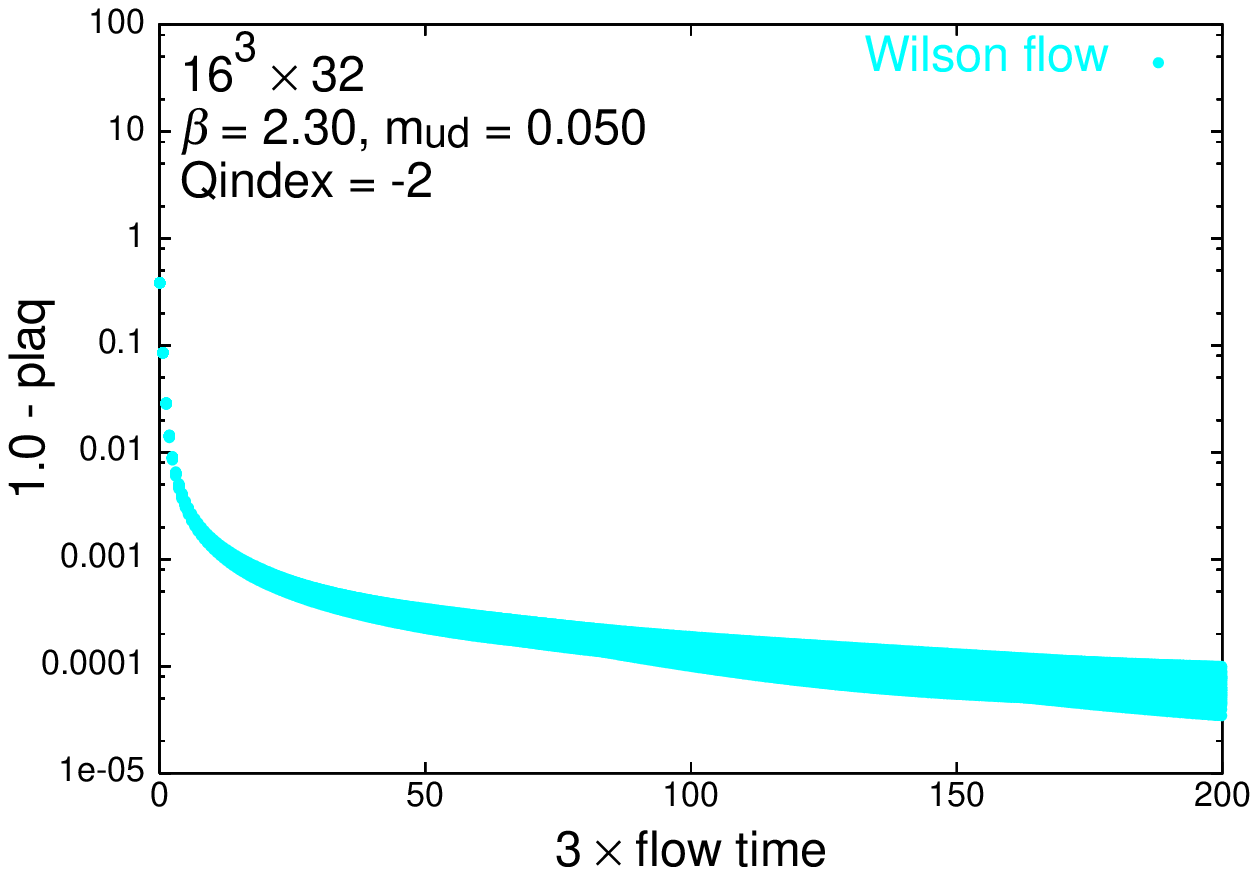}
 \caption{
   Wilson flow time dependence of
   maximum of $(1 - \mbox{plaquette})$ (left panel),
   and the value summed over the spacetime volume (right panel).
 }
 \label{figure:wilson_flow_time_vs_plaq}
\end{center}
\end{figure}

Figure~\ref{figure:cooling_vs_topological_charge} illustrates
cooling and smearing step dependence of the improved topological charge.
The flow time dependence is also plotted.
The flow time is multiplied by a factor of three,
which is expected from a perturbative analysis~\cite{Bonati_2014}.
In every case,
the topological charge has an integer value
with a sufficiently large number of steps.
%
%
%
A small number of smoothing steps
leads to a fake plateau i.e. a semi-stable value of
the topological charge.
%
%
%
%
The number of smoothing steps is determined
to satisfy the admissibility condition,
$\max [ \mbox{Re} \mbox{Tr} (1 - U_{\rm plaq}) ] < 0.067$~\cite{Luescher_1982_2010,Phillips_1986}.
In Fig.~\ref{figure:wilson_flow_time_vs_plaq},
Wilson flow time dependence of the plaquette
is shown.
%
%
No jump of the topological charge seem to be triggered,
if the admissibility condition is fulfilled.
It should be mentioned $\max [ \mbox{Re} \mbox{Tr} (1 - U_{\rm plaq}) ]$ does not
always decrease as the flow time grows,
though the value summed over the spacetime falls off monotonically.

\begin{figure}[t]
\begin{center}
 \includegraphics[width=45mm]{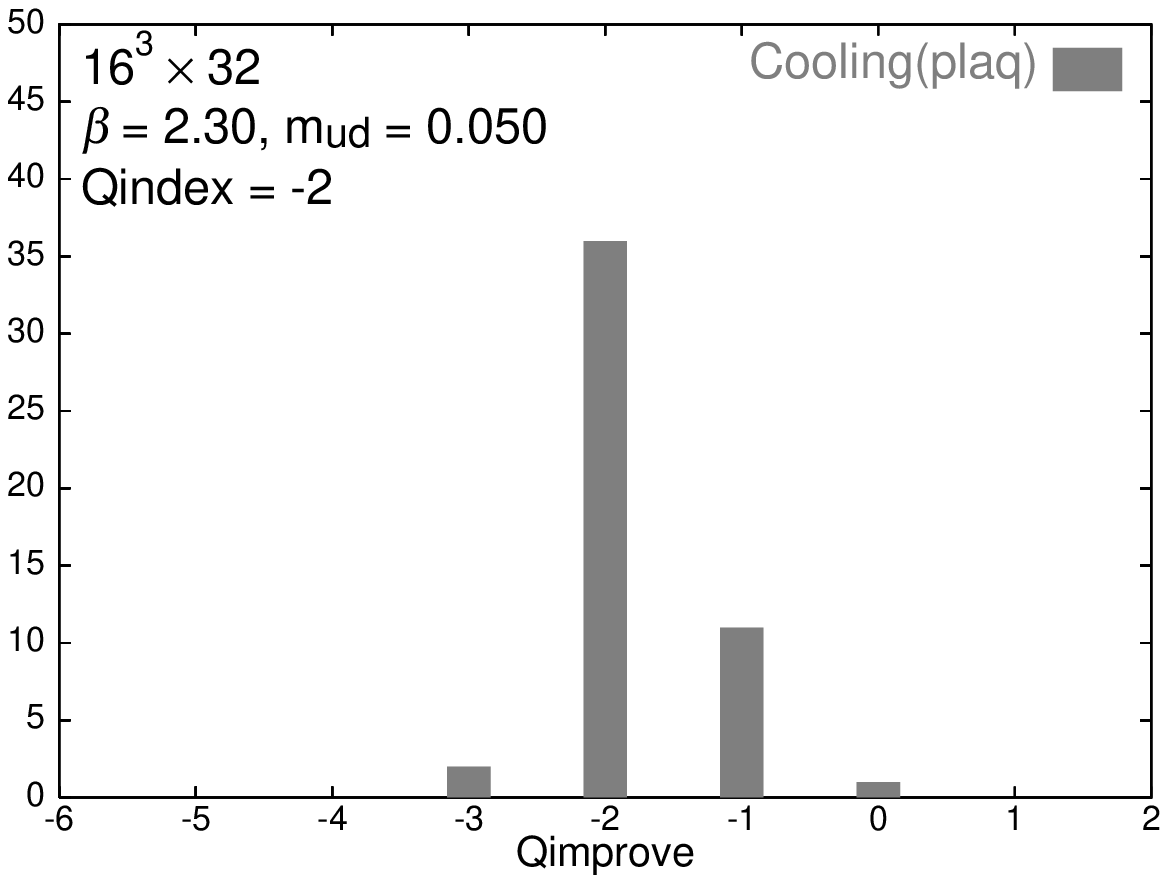}
 \includegraphics[width=45mm]{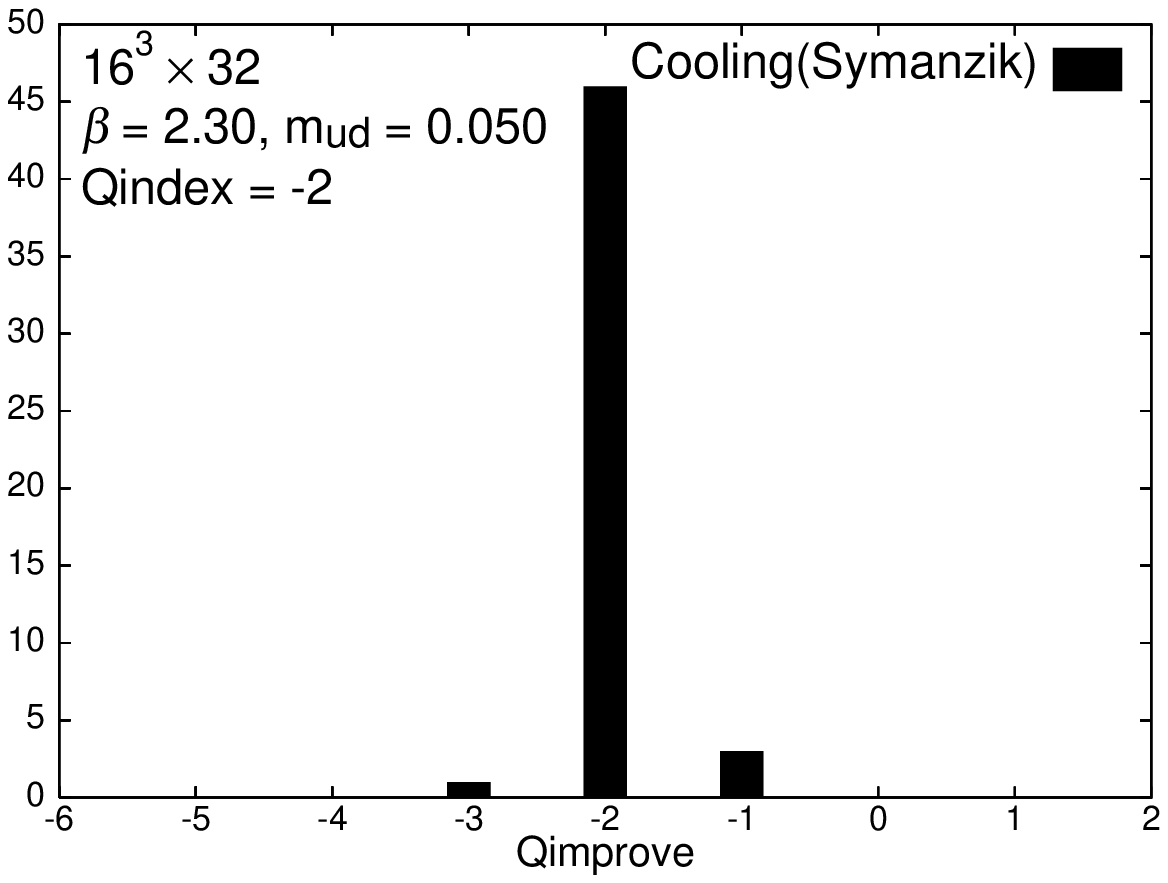}
 \includegraphics[width=45mm]{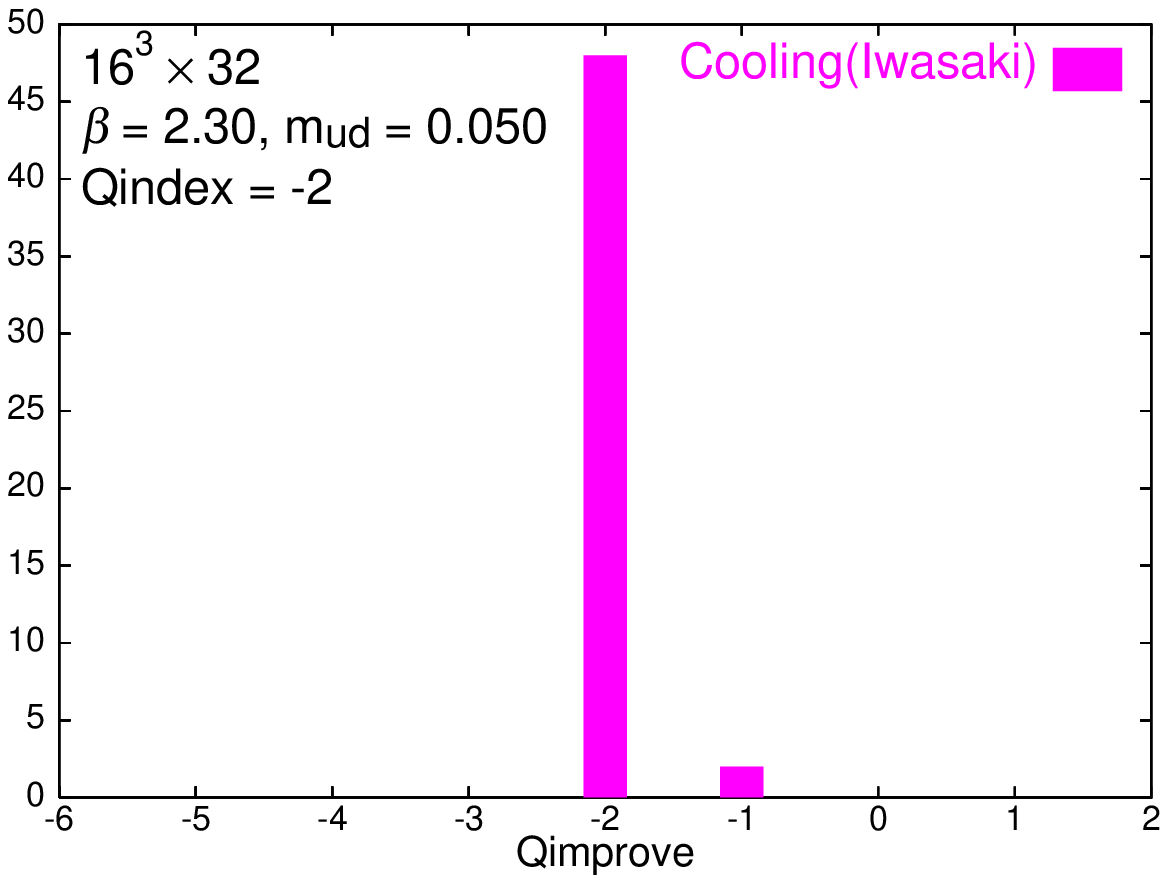}
 \\
 \includegraphics[width=45mm]{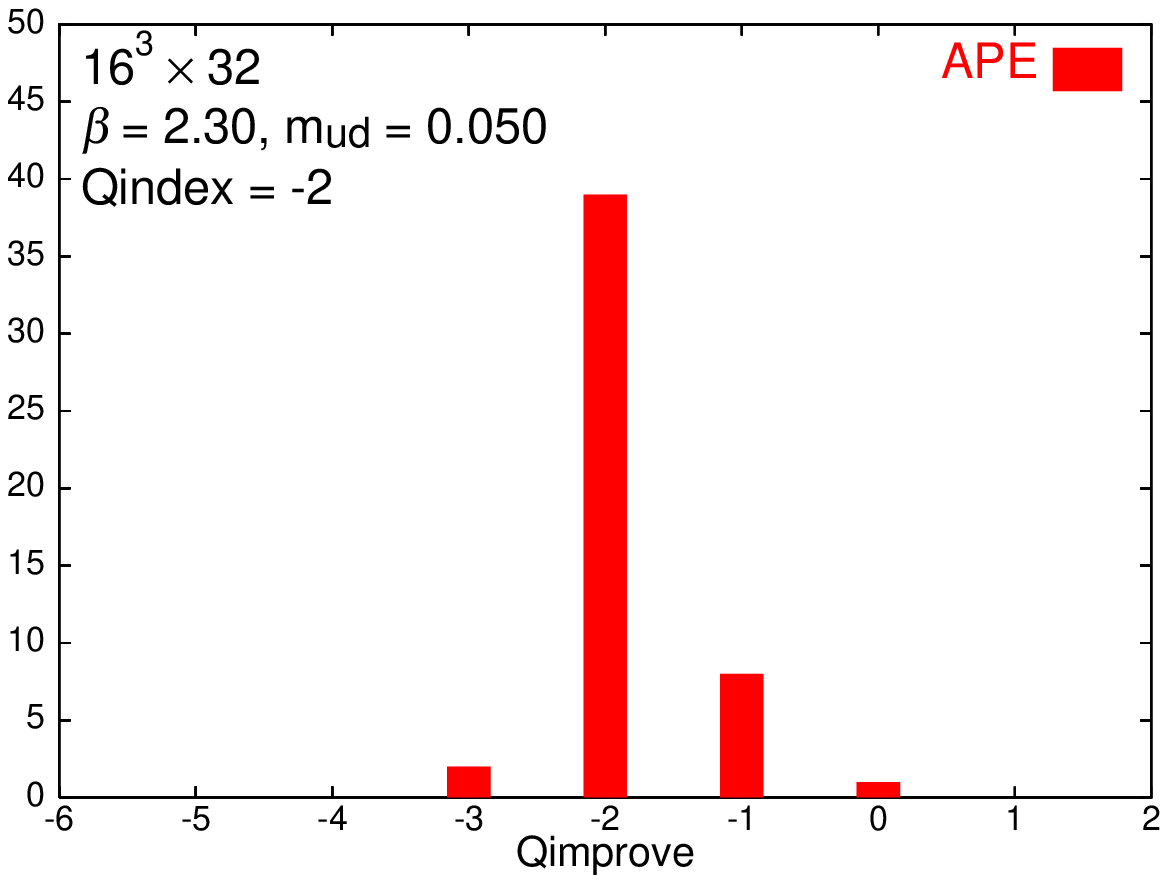}
 \includegraphics[width=45mm]{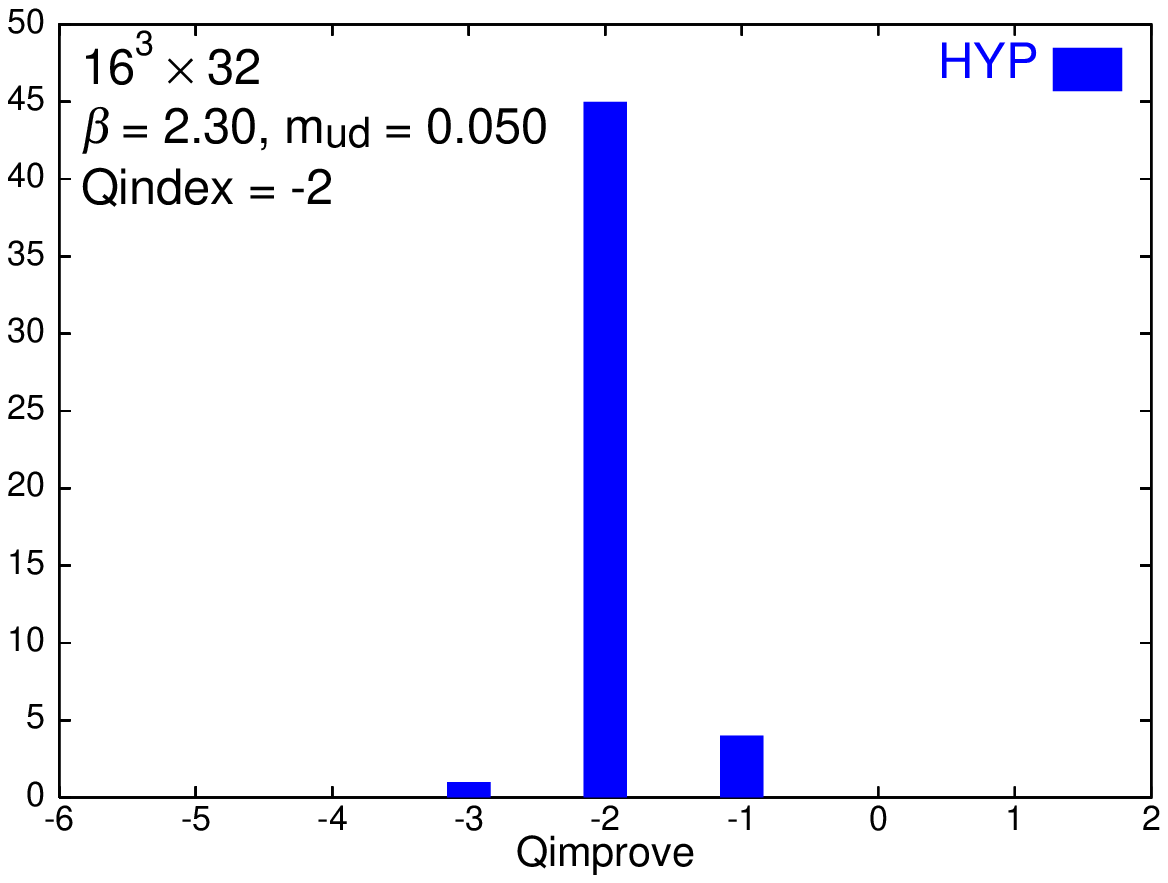}
 \\
 \includegraphics[width=45mm]{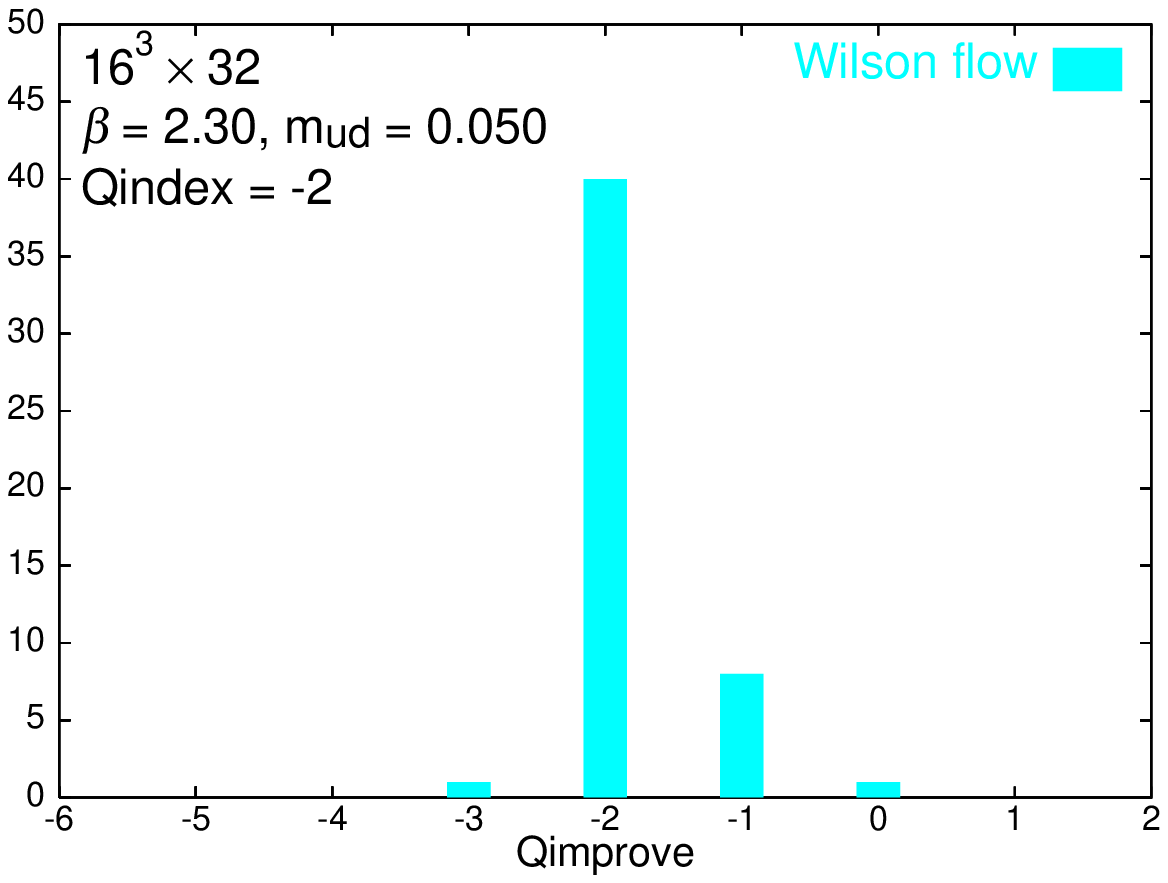}
 \includegraphics[width=45mm]{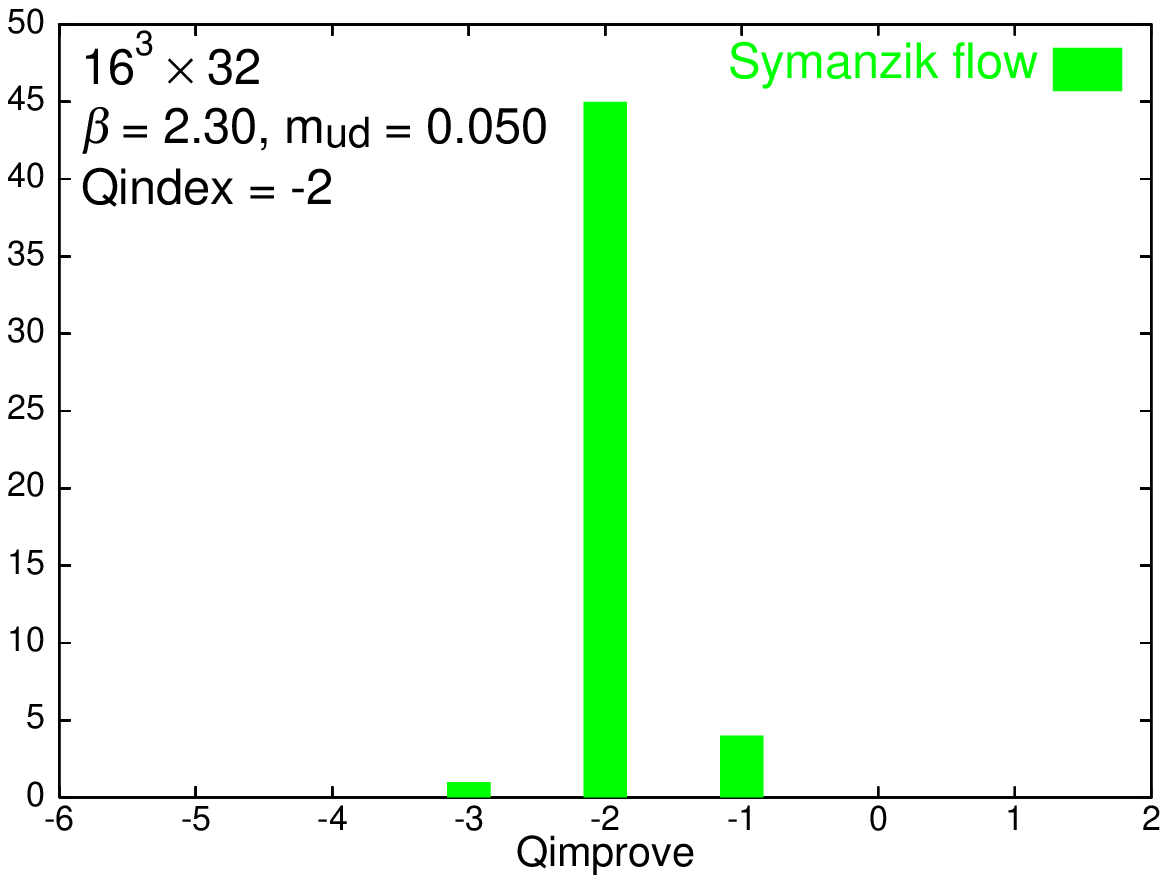}
 \includegraphics[width=45mm]{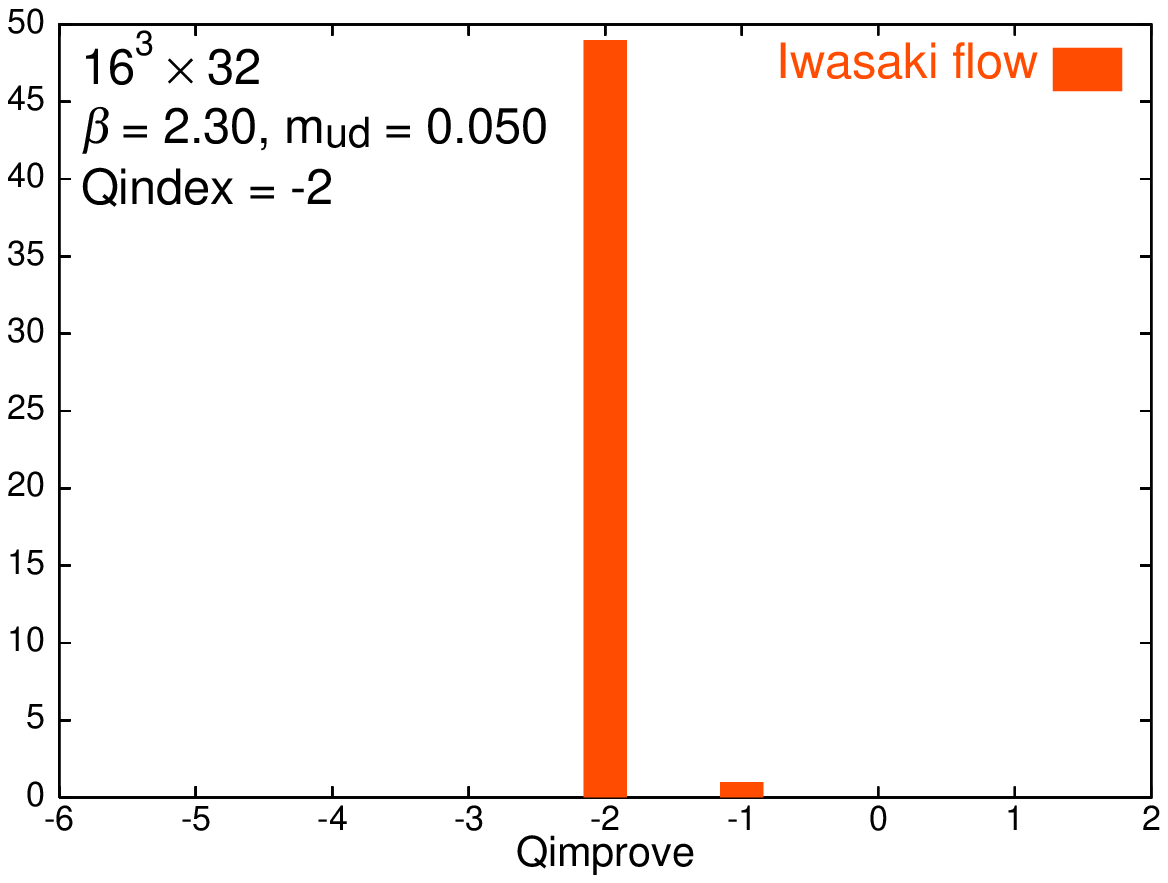}
 \caption{
   Histograms of $Q_{\rm improve}$
   with cooling (top panels),
   smearing (middle panels),
   and gradient flow (bottom panels).
 }
 \label{figure:Q_histogram_on_Qindex_minus_2}
\end{center}
\end{figure}

Figure~\ref{figure:Q_histogram_on_Qindex_minus_2} presents
histograms of the improved topological charges.
Since the topological charge determined by the index $Q_{\rm index}$
is fixed in the configuration generations,
the histogram is expected to have a sharp peak around $Q_{\rm index}$,
supposing the scaling violation is small.
The histograms obtained by cooling with improved local actions
show the expected behavior.
Almost all of the topological charges agree with $Q_{\rm index}$.
On the other hand, cooling using the plaquette action has a broad histogram.
It implies a relatively large lattice artifact in the unimproved cooling method.
Analogous trends are observed in other smoothing procedures.
HYP smearing has a narrow histogram, while APE smearing does not.
Symanzik and Iwasaki flows form a sharp peak in the histogram.
On the contrary, Wilson flow brings a wide peak.
Improved smoothing methods leads to higher consistency with $Q_{\rm index}$,
indicating the scaling violation in the topological charge is suppressed well by the improvement.


\subsection{Conclusion}

Systematic comparison of topological charges is presented.
Topological charges are measured on $N_f=2$ topology fixed configurations.
%
Several smoothing techniques are evaluated using a gluonic topological charge operator
of Symanzik-type coefficients,
which give a topological charge with the smallest deviation from an integer.

Cooling with improved actions, HYP smearing, and improved gradient flows
are found to be advantageous.
More than 90\% of the topological charges are consistent with
those obtained by the index theorem.
It indicates their lattice artifacts are reduced efficiently.
On the other hand,
cooling with plaquette action, APE smearing, and Wilson flow
lead to partial matches.
The agreement is limited to 70-80\%.
Scaling violations seem to be comparatively large in these smoothing methods.

Scaling properties as well as finite size effects of the topological charge
have not been investigated.
It is important to estimate them, but is beyond the scope of this work
due to limitation of the gauge configurations.
The gauge configurations have been generated
at a single lattice spacing and spatial volume.
These evaluations are left for the future work.

\section*{Acknowledgments}

I would like to thank
members of MEXT SPIRE Field 5 Project 1
and Bridge++ development team,
as well as H.Fukaya and K-I.Nagai
for valuable discussions.
I am grateful to JLQCD Collaboration
for providing their gauge configurations~\cite{JLQCD_2008}.
This work is in part based on Bridge++ code~\cite{Bridge,Bridge_ueda,Bridge_motoki}.
This work is supported by JLDG constructed over SINET3 of NII,
MEXT SPIRE and JICFuS,
and Grants-in-Aid for Scientific Research Grant Number 24540250.



\begin{thebibliography}{99}
 \bibitem{Plenary_2014}
 For a recent review, see
 M.~Mueller-Preussker,
 PoS (LATTICE2014), 003 (2014).
 %
 %
 %
 \bibitem{Talk_2014}
 K.~Cichy et al.,
 PoS (LATTICE2014), 017 (2014);
 P.~Giudice et al.,
 PoS (LATTICE2014), 273 (2014).
 %
 \bibitem{JLQCD_2008}
 JLQCD Collaboration: S.~Aoki et al., 
 Phys. Rev. D78, 014508 (2008).
 %
 \bibitem{Cooling_1981}
 B.~Berg,
 Phys. Lett. 104B, 475 (1981);
 M.~Teper,
 Phys. Lett. 162B, 357 (1985);
 J.~Hoek et al.,
 Rutherford preprint RAL-85-042 (1985);
 E.~M.~Ilgenfritz et al.,
 Nucl. Phys. B268, 693 (1986).
 %
 \bibitem{APE_1987}
 APE Collaboration: M.~Albanese et al.,
 Phys. Lett. B192, 163 (1987).
 %
 \bibitem{HYP_2001}
 A.~Hasenfratz and F.~Knechtli,
 Phys. Rev. D64, 034504 (2001).
 %
 \bibitem{Luescher_2010}
 M.~L\"{u}scher,
 Commun. Math. Phys. 293, 899 (2010).
 %
 \bibitem{Celledoni_2003}
 E.~Celledoni et al.,
 FGCS 19, 341 (2003).
 %
 \bibitem{Bonati_2014}
 C.~Bonati and M. D'Elia,
 Phys. Rev. D89, 105005 (2014).
 %
 \bibitem{Luescher_1982_2010}
 M.~L\"{u}scher,
 Commun. Math. Phys. 85, 39 (1982);
 ibid. JHEP 08, 071 (2010).
 %
 \bibitem{Phillips_1986}
 A.~Phillips and D. Stone,
 Commun. Math. Phys. 103, 599 (1986).
 %
 \bibitem{Bridge}
 http://bridge.kek.jp/Lattice-code/
 %
 \bibitem{Bridge_ueda}
 S.~Ueda et al.,
 PoS (LATTICE2013), 412 (2014);
 PoS (LATTICE2014), 036 (2014);
 ibid. J. Phys. Conf. Ser. 523, 012046 (2014).
 %
 \bibitem{Bridge_motoki}
 S.~Motoki et al.,
 Proc. Comp. Sci. 29, 1701 (2014).
\end{thebibliography}
\end{document}